# Securing Cloud from Cloud Drain

Niva Das[*]
*Corresponding Author*
Calcutta University
Kolkata, India
niva.cu@gmail.com

Tanmoy Sarkar
Neudesic India Pvt. Limited
Hyderabad, India
tanmoy.sarkar@neudesic.com

*Abstract*— **Today smart devices are growing rapidly with the rising demand of communication among themselves. Internet of Things (IoT) revolutionized the way of communicating among device and in this revolution cloud computing plays a major role. To identifying a device anywhere in the world and communicating with it is a major challenge. So, cloud computing [1] comes for the rescue. Many companies are investing a large amount of resources to provide services to customer through cloud. For e.g. google releases Google Cloud Messaging to update devices automatically from the server using device id. Also, we can achieve a lot from cloud services like creating VMs, Storages as per need. With the increasing usage of cloud increases cloud security concern. In this paper, we will discuss about cloud security methods. We have used the term "Cloud-Drain" to define data leakage in case of security compromise.**

## I. Introduction And services provided

In recent years cloud computing is perhaps the most discussed topic as more and more companies are trying to adapt and provide this service to its partners/clients. The advantages of cloud computing is reduced maintenance, cost efficiency, redundancy of data, scalability, increase in storage. There are some disadvantages also like vendor lock-in, downtime, limited control and increased vulnerability. Cloud computing mainly provides three models:

Infrastructure as a Service (IaaS) allows user to provision their infrastructure resources to cloud. Platform as a Service (PaaS) allows cloud providers to deliver the user development environment services where user can develop and run applications which are built in-house. Software as a Service (SaaS) [13] allows the cloud provider to deliver the capability to run their application on cloud. The applications are accessible through thin client such as web browser Fig 1.

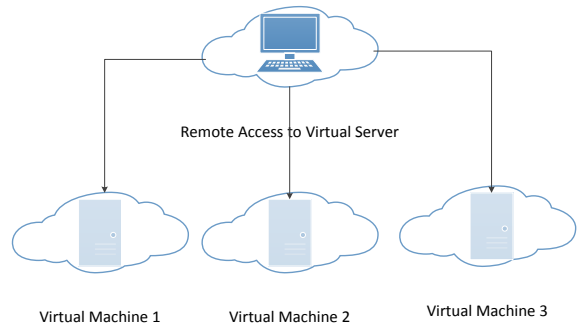

Figure 1

Cloud computing technology simplifies infrastructure planning and provides business-owner the opportunity to deploy applications on private, public, hybrid or community clouds, based on their needs. With this advancement of cloud where the customers can choose the type of cloud suits, there business increases but there is increase of security concerns and data leakage. Customers uploaded there highly confidential data which if compromised will drastically affect their business. Private cloud is always an option where customer can manage their own security policies and their resources. They will however need to access the cloud using public networks. There are many papers [2] [3] which do survey on cloud security, attacks and its mitigation.

In this paper we are using a new term "cloud drain" to acknowledge the security issues in cloud computing and also discuss few of the mitigation problems.

## II. literature review

Some methods have been suggested in literature for handling security issues in organizations implementing cloud computing

Maggi and Zanero [16] addressed countermeasures (anti-viruses, intrusion detection systems), developed to mitigate well-known security threats. The focus is mainly on anomaly-based approaches which are mostly suited for modern protection tools and not for earlier generation intrusion detectors. The pattern-based changes (example: from thin client connected to the main frame or powerful workstations connecting to thin clients) are observed, which cause some simultaneous changes in work environment and generates new problems to security of Cloud Computing.

Ertaul et al. [17] discussed cloud computing features like reduced total cost of ownership and scalability issues. They claim that cloud computing also minimizes complexity and provides service to customers. Virtualization is the technique used to deal with quality of service (QOS). Usage of cloud computing is considered to be unsafe in an organization. For dealing with this type of situation, they investigated a few major security issues with cloud computing and also the existing countermeasures to those security challenges. Advantages for implementing cloud computing from a different point of view are also discussed.

Subashini and Kavitha [18] dealt with the security risks faced in cloud computing. They provided empirical evidence on security risks and issues encountered during deployment of service delivery models in an organization. The service models are placed in cloud and the empirical validation was made in order to justify the safety of the environment. Security was the main issue while there were also complications with data protection and data privacy in a continuous manner that affected the market.

Md. Tanzim Khorshed et al [19] boast that cloud computing helps reduce cost of services and improves business outcomes. But to market this and popularize its use by IT user community, there are many security risks to be solved. They also mentioned that cloud services pose an attractive target for cyber-attacks and criminal activities. This is due to the fact that these services have sensitive information from many organizations and individuals, stored in their repositories.

Iliana Iankoulova et al [20] have performed a systematic review to identify which security requirements need to be further researched. From this review they found that nonrepudiation, physical protection, recovery and prosecution are the least researched in security areas. Integrity, access control and security auditing are the most popular addition to security requirements; solutions to these identified challenges were also mentioned. Vasudevan et al. [9] described a novel multipath approach of data communication to improve security; this will be relevant to transfer of data to and fro from the cloud. Christodorescu et al. [11] provided a thorough analysis of cloud security and provided how it is different from virtualization security issues.

## III. CLOUD DRAIN

The term cloud-drain is used to signify the data leakage in case of security compromise in cloud computing. Intrusion plays a major role in this part. An intrusion consists of an attack exploiting a security flaw and a consequent breach which is the resulting violation of the explicit or implicit security policy of the system. Here we discuss about intrusion techniques and how they will effectively allow cloud drain possible. Intrusion Detection techniques could be used to prevent a great deal of intrusion which affects the stored data in the cloud. [21] [22] and [24]. Neural networks and application in Cloud Security is also a well-researched area [23].

Cloud systems are susceptible to all typical network and computer security attacks. The targets that are possibly vulnerable are the protocol stack; network devices; processes running in kernel space, such as operating system daemons; and processes running outside kernel space, such as cloud middleware, cloud applications, and any non-cloud applications running with either root or user privileges. Classification of cloud intrusions is given as follows:

*1) Unauthorized Access:* A break-in committed by an intruder who can pose as a legitimate cloud user if the credentials is obtained by stealing, brute force or careless disclosure, then it can cause devastating results. Man-in-the- middle attack can also be a possibility to gain illegitimate user access. During 2010, only 4 million user accounts were compromised by hackers; in 2011 hackers penetrated 174 million accounts (thanks, Anonymous), according to the Data Breach Investigations Report published by Verizon in March[10].

*2) Network Attack:* Attacks performed with the help of tools or exploit scripts that target vulnerabilities existent in cloud protocols, services and applications. They may appear in the form of DOS attacks, probes, and worms, and may leave their traces at several locations of cloud's organization.

*3) Data Security:* Data in cloud is stored in different geographical locations. Cloud providers are providing maximum security to these places. But since the consumers are unaware of physical locations, data breach is always a possible and valuable data may disappear without a trace. A malicious user of a Virtual Machine (VM) or careless cloud provider could be responsible for data loss. Also, natural disaster can cause data loss if data is not properly distributed. Data duplication in a remote cloud at periodic interval is a possible solution to mitigate such issues.

The Cloud Security Alliance found the below few mentioned threats that can lead to cloud drain:
Failures in Provider Security, Attacks by Customers, Unavailability and Untrustworthiness Issues, Legal and Regulatory Issues, Data Loss, Account hack.

All these threat can lead to cloud drain and drastically affect the cloud consumers. Fundamentally the cloud security model involves of cloud providers, service providers and cloud consumers. Each entity has their own security policies and management. So, before engaging to cloud services, all the three entities need to negotiate and come to an agreement. The security gap can be explored in more detail and leads to security attacks which in turn may lead to cloud drain.

## IV. CLOUD DRAIN PREVENTION

Securing VM's operating system, repositories and network can prevent cloud drain in physical servers. Many tenants VM's are sharing the same infrastructure and may lead to security vulnerabilities. One potential risk has to do with the potential to compromise a virtual machine. Virtual machine compromise can cause major impact and destruction. This requires an additional degree of network isolation and enhanced detection by security monitoring. For this, hypervisors are used because at present there are no documented attacks against hypervisors, reducing the likelihood of attack. So, the vulnerability of the hypervisor and the probability of an attack are low [4].

Data security [5] is another important aspect in cloud computing. Many organizations [7] and researchers [6] [8] have implemented encryption for data security; but they often overlooked inherent weaknesses in key management and data access.

If encryption keys are not protected, they are susceptible to theft by malevolent hackers. Vulnerability also lies in the access control model; thus, if keys are appropriately protected but access is not sufficiently controlled or robust; malicious or compromised personnel can attempt to access sensitive data by assuming the identity of an authorized user.

First, we need to make sure that data is not readable and that the solution offers strong key management. Second, implement access policies that ensure only authorized users can gain access to sensitive information, so that even privileged users such as root user cannot view sensitive information [15]. Third, incorporate security intelligence that generates log information, which can be used for behavioral analysis to provide alerts that trigger when users are performing actions outside of the norm.

Use advance automatic switches which provides packet rate inspection and bogus IP filtering (Bogon filtering). Intelligent hardware that is Application Front end hardware device is placed on the network which analyzes the data packets entering the network system and identifies that whether they are based on priority, regular or dangerous. DoS attacks [12] and Malware-Injection Attack Solution is to deploy a Hypervisor in the provider's side. This Hypervisor will be considered the mainly secured and complicated part of the cloud system whose protection cannot be breached by any means. The Hypervisor is responsible for arrangement of all the Instances. It cannot be denied that there are risks associated with sharing the same physical infrastructure between a set of multiple users, even one being malicious can cause threats to the others using the same infrastructure [14], and hence security with respect to hypervisor is of great concern as all the guest systems are controlled by it for avoiding a flooding attack; our projected approach is to categorize all the servers in the cloud system as a cluster of fleet of servers. Each fleet of servers will be elected for particular type of job, in this; all servers in the fleet will have inside communication along with themselves through message passing. So when a server is loaded, a latest server will be installed in the fleet and the name server, which has the entire records of the existing states of the servers, will renew the target for the requests with the latest included server.[10]

## V. CONCLUSION

Although security researches are done in many fields like Ad-hoc networks [25] [26], mobile computing [27], sensor networks [28], radio networks etc. but cloud computing is always exposed to security threats varying from network level threats to application level threats. In order to keep the Cloud secure, these security threats need to be controlled. Moreover data residing in the cloud is also prone to a number of threats and various issues like: confidentiality and integrity of data should be considered while buying storage services from a cloud service provider. Inspecting of the Cloud regularly needs to be done to safeguard the cloud against external threats and avoid cloud drain